\documentclass[aip,pop,amsmath,amssymb,showpacs,reprint,floatfix,lengthcheck]{revtex4-1}
\usepackage[colorlinks,linkcolor=blue,citecolor=blue,bookmarks,pdfstartview=FitH]{hyperref}
\usepackage{graphicx}
\usepackage{dcolumn}
\usepackage{bm}
\usepackage{url}
\usepackage{amssymb}
\usepackage{enumerate}
\usepackage{enumitem}

\begin{document}

\preprint{AIP/123-QED}

\title{Verification of particle simulation of radio frequency waves in fusion plasmas}

\author{Animesh Kuley}
\email{akuley@uci.edu}
\affiliation{Department of Physics and Astronomy, University of California Irvine, CA 92697,USA}
\affiliation{Fusion Simulation Center, Peking University, Beijing 100871, China}
\author{Z. X. Wang}
\affiliation{Department of Physics and Astronomy, University of California Irvine, CA 92697,USA}
\author{Z. Lin}
\affiliation{Department of Physics and Astronomy, University of California Irvine, CA 92697,USA}
\affiliation{Fusion Simulation Center, Peking University, Beijing 100871, China}
\author{F. Wessel}
\affiliation{Tri Alpha Energy, Inc., Post Office Box 7010, Rancho Santa Margarita, California 92688, USA}

\date{\today}

\begin{abstract}
Radio frequency (RF) waves can provide heating, current and flow drive, as
well as instability control for steady state operations of fusion
experiments. A particle simulation model has been developed in this work to
provide a first-principles tool for studying the RF nonlinear interactions
with plasmas. In this model, ions are considered as fully kinetic particles
using the Vlasov equation and electrons are treated as guiding centers using the drift kinetic equation. This model has been implemented
in a global gyrokinetic toroidal code (GTC) using real
electron-to-ion mass ratio. To verify the model, linear simulations of ion plasma oscillation, ion Bernstein wave, and lower hybrid wave are carried out in cylindrical geometry and found to agree well with analytic predictions.
\end{abstract}

\maketitle

\section{Introduction}
The importance of radio frequency (RF) waves as a source for heating and current drive has been recognized from the early days of magnetically confined plasma research\cite{RevModPhys.59.175,Gormezano2007}. The RF waves provide one of the very few options for steady state operation of the burning plasma experiment ITER, the crucial next step in the quest for the fusion energy. First, the RF waves in ITER will be used to deliver sufficient central heating power to access the H-mode confinement regime and to control the plasma temperature. Secondly they can provide a non-inductive central current drive and an off-axis current drive capability for the current profile control. Thirdly they will be used for the control of magnetohydrodynamic (MHD) instabilities in ITER. It has also been proposed\cite{Rostocker} that the RF waves can be used for driving plasma flows and current in the field reversed configuration\cite{PhysRevLett.105.045003}. To effectively utilize the RF power we need a better understanding of the key physics of RF waves in plasmas, e.g., wave-particle interaction\cite{Kominis, Lee,Hellsten}, mode conversion\cite{Y2011,Tsujii} and nonlinear effects\cite{Tripathi1979,liu1984density,kuley2010lower,Animesh2011,Cesario2011,Goniche,gao2011radial}

Two computational methods have been widely used to study wave-particle interactions in fusion plasmas. The first solves the wave equation derived from the linearized Vlasov-Maxwell system (the full wave model).  This approach has been used in the eigenvalue solvers like TORIC\cite{Brambilla} and AORSA\cite{Jaeger2001} to study high frequency waves such as the lower hybrid wave and the ion Bernstein wave.  However, this method does not capture the crucial nonlinear physics.  The second method is the initial value simulation in which a kinetic equation is integrated in time, retaining all nonlinearities.  Such an approach has been taken by gyrokinetic (GK) simulation codes, which have revolutionized studies of turbulent transport driven by low frequency drift waves\cite{Lee1983,Lin98}. Nonlinear phenomena of the RF waves have been studied in the slab geometries with particle codes such as GeFi\cite{qi2013simulation}, Vorpal\cite{Jenkins} and G-gauge\cite{Zhi}.

For waves in the intermediate frequency range, between the ion and electron cyclotron frequencies (e.g., lower hybrid wave, ion Bernstein wave, etc.), the GK model is not valid, but a fully kinetic model for both ions and electrons is inefficient due to the small electron-to-ion mass ratio. These waves often play important roles in the kinetic processes of magnetized plasmas, e.g., particle acceleration, current drive, plasma heating and spectral cascade of turbulence from long to short wavelength. In this work, we develop a simulation model for these waves, which uses fully kinetic (FK) ions but treats electrons in the drift kinetic approximation (DK). We will study only waves with wavelength longer than the electron gyroradius, so that the electron GK equation reduces to the DK equation. The current FK/DK hybrid simulation model\cite{chen09}  can be regarded as a reduced version of the FK/GK model\cite{Yu}, which overcomes the difficulty associated with the small electron mass by analytically removing the high frequency modes (electron cyclotron frequency and electron plasma frequency). Our goal is to develop a new nonlinear toroidal particle simulation model, which is the most effective approach to study the nonlinear physics in the RF heating and current drive.

Realistic RF simulations for fusion plasmas also require the global toroidal geometry and massively parallel computing due to multiple temporal and spatial scales. The current work utilizes the gyrokinetic toroidal code (GTC)\cite{Lin98} to take advantage of its existing physics capability, toroidal geometry and computational power. GTC has been extensively applied to study turbulent transport in fusion plasmas including ion and electron temperature gradient turbulence,\cite{Lin1999,Lin2002,Lin2007} collisionless trapped electron mode turbulence,\cite{Xiao2009}  energetic particle turbulence and transport\cite{Zhang2008,Zhang2011,Huassen2012,Deng2012} and neoclassical transport\cite{Lin1997}. As a first step in developing this nonlinear toroidal particle simulation model, the verification of the linear physics of lower hybrid   wave (LHW) and ion Bernstein wave (IBW)  in cylindrical geometry are presented in this paper. 

The paper is organized as follows: the fully kinetic ion and drift kinetic electron simulation model is described in Sec. II, Sec III gives the verification of the GTC simulation of the electrostatic normal modes in uniform plasmas. Sec. IV summarizes this work.

\section{Formulation of fully Kinetic ion  and Drift kinetic electron simulation model}
\subsection{Formulation of FK ion and DK electron model}
The FK ion and DK electron simulation model treats the ion with the fully kinetic (FK) model and the electron with the drift kinetic (DK) approximation. For the FK ion, the dynamics is described by the six dimensional Vlasov equation
\begin{equation}
\biggl[\frac{\partial}{\partial t}+\dot{\textbf{x}}\cdot \nabla+\frac{Z_i}{m_i}(\textbf{E}+\textbf{v}\times \textbf{B}_0)\cdot\frac{\partial}{\partial \textbf{v}}\biggr]f_i=0,
\end{equation}
where $f_i$ is the ion distribution function, $Z_i$ is the ion charge and $m_i$ is the ion mass. $\textbf {B}_0=B_0\textbf{b}_0$ is the equilibrium magnetic field. In the current simulation we use the cylindrical coordinates $\textbf{x}(r,\theta,\zeta)$, where $r$ is the radial position, $\theta$ is the poloidal angle and $\zeta$ is the length of the cylinder with circular cross section. The evolution of the ion distribution function $f_i$ can be described by the Newtonian equation of motion in the presence of self-consistent electromagnetic field as follows 
\begin{eqnarray}
\frac{d\textbf{x}}{dt}=\textbf{v}_{\perp}+\textbf{b}_0 v_{\parallel},\nonumber\\
\frac{dv_{\parallel}}{dt}=\frac{Z_i}{m_i}\textbf{b}_0\cdot \textbf{E},\\
\frac{d\textbf{v}_{\perp}}{dt}=\frac{Z_i}{m_i}(\textbf{E}_{\perp}+\textbf{v}_{\perp}\times \textbf {B}_0).\nonumber
\end{eqnarray}
In the fully kinetic version of the GTC code we use $\textbf{v}(v_{\parallel},v_{\perp},\alpha)$ for the velocity space, where $v_{\parallel}$ and $v_{\perp}$ are the parallel and perpendicular velocity, respectively, and $\alpha$ is the gyro phase angle. This model retains full finite Larmor radius effects and wave frequencies larger than $\omega_{ci}$, where $\omega_{ci}$ is the ion gyrofrequency.

Electron dynamics is described by the drift kinetic equation using guiding center position $\textbf{X}(r,\theta,\zeta)$, perpendicular $(v_{\perp})$ and parallel $(v_{\parallel})$ velocity  as a set of independent variables 
\begin{equation}
\biggl[\frac{\partial}{\partial t}+\dot{\textbf{X}}\cdot \nabla+\dot{v}_{\parallel}\frac{\partial}{\partial v_{\parallel}}\biggr]f_e=0,
\end{equation}
where $f_e$ is the guiding center distribution function. The evolution of the electron distribution function can be described by the following equations of guiding center motion\cite{Brizard2007}
\begin{eqnarray}
\frac{d\textbf{X}}{dt}=\textbf{v}_{E}+\textbf{b}_0 v_{\parallel},\nonumber\\
\frac{dv_{\parallel}}{dt}=-\frac{e}{m_e}\textbf{b}_0\cdot \textbf{E},
\end{eqnarray}
where $dv_{\perp}/dt=0$ (by definition), $\textbf{v}_E=(\textbf{E}\times \textbf{b}_0)/B_0$. The above Eq.(4) is valid only for uniform magnetic field. This electron model is suitable for the dynamics with the wave frequency $\omega <\omega_{ce}$ and $k_{\perp}\rho_e \ll 1$, where $k_{\perp}$ is perpendicular to the magnetic field, $\omega_{ce}$ is the electron cyclotron frequency and $\rho_e$ is the electron gyroradius.

The electrostatic potential $\phi$ can be found from the Poisson's equation
\begin{equation}
\biggl(1+\frac{\omega_{pe}^2}{\omega_{ce}^2}\biggr)\nabla_{\perp}^2\phi=-4\pi(Z_in_i-en_e),
\end{equation}
assuming $|\nabla_{\perp}^2| \gg |\nabla_{\parallel}^2| $ to suppress the undesirable high frequency electron plasma oscillation along the magnetic field line. Second term on the left hand side corresponds to the electron density due to its perpendicular polarization drift of the electrostatic field. The number densities are defined as the fluid moments of the corresponding distribution function,
\begin{eqnarray}	
		n_i=\int dv_{\parallel}v_{\perp}dv_{\perp}d\alpha f_i, \nonumber\\
		n_e=2\pi\int dv_{\parallel}v_{\perp}dv_{\perp} f_e.
\end{eqnarray}

 Eqs. (2)-(6) are implemented using both non-perturbative (full-$f)$ and perturbative $(\delta f)$ methods in GTC. We use the $\delta f$ simulation for the fully kinetic ion to reduce the particle noise in this work. In the current linear simulation, we assume that the background plasma is uniform in density and temperature. We decompose the ion distribution function into its equilibrium $f_{0i}$ and perturbed part $\delta f_{i}$, where $(\delta f_{i}\ll f_{0i})$. By defining the particle weight $w_{i}=\delta f_i/f_{0i}$ for the linear simulation, we can rewrite the Vlasov equation for ion as follows

\begin{figure}[bpt]
	\centering
	\includegraphics{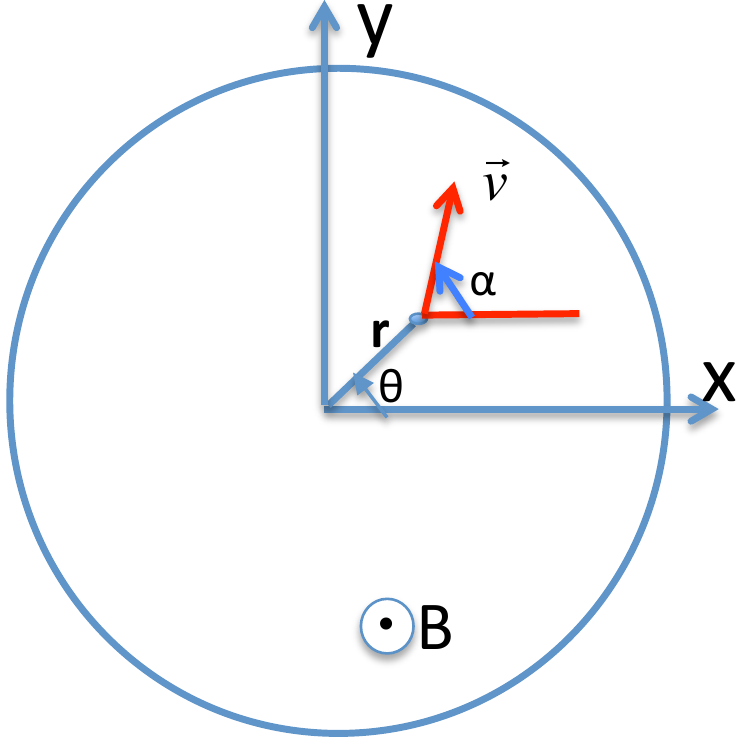}
	\caption{Coordinate system on the poloidal cross section of a cylinder.}
\end{figure}

\begin{eqnarray}
\frac{dw_i}{dt}=-\frac{1}{f_{0i}}\biggl[\frac{Z_i}{m_i}E_{\parallel}\frac{\partial}{\partial v_{\parallel}}+\frac{Z_i}{m_i}\textbf{E}\cdot \frac{\partial}{\partial{\textbf{v}_\perp}}\nonumber\\
+\frac{Z_i}{m_i}\frac{\textbf{E}\cdot(\hat{b}_0\times \textbf{v}_{\perp})}{v_{\perp}^2}\frac{\partial}{\partial\alpha}\biggr]f_{0i},
\end{eqnarray}
where the second and third terms on the right hand side arise due to the change in the perpendicular energy and the correction of the gyro frequency, respectively. By considering the background plasma as a Maxwellian with the temperature $T_i$, one can further simplify the weight equation as follows
\begin{equation}
\frac{dw_i}{dt}=\biggl[\frac{Z_i}{T_i}E_{\parallel}v_{\parallel}+\frac{Z_i}{T_i}\textbf{E}\cdot \textbf{v}_\perp\biggr]
\end{equation}
Similarly the weight equation for the electron in a uniform Maxwellian background with the temperature $T_e$ can be written as\cite{Holod}
\begin{equation}
\frac{dw_e}{dt}=-\frac{e}{T_e}E_{\parallel}v_{\parallel},
\end{equation}
where $w_e=\delta f_e/f_{0e}$ for the linear simulation. $f_{0e}$ and $\delta f_e$ are the equilibrium and perturbed distribution function, respectively. Eqs. (8) and (9) are valid only for uniform density and temperature. The parallel component of the electric field can be written as 
\begin{equation}
E_{\parallel}=-\textbf{b}_0\cdot\nabla\phi
\end{equation}
With a uniform magnetic field one can write down the change in the perpendicular energy as follows
\begin{equation}
\textbf{E}\cdot \textbf{v}_\perp=-\dot{\theta}\frac{\partial \phi}{\partial\theta}-\dot{r}\frac{\partial \phi}{\partial r},
\end{equation}
where the particle equations of motion in cylindrical coordinates are
\begin{eqnarray}
\dot{\zeta} = \frac{v_\parallel}{R_0},\nonumber\\
\dot{\theta} = \frac{v_\perp}{r}\text{sin}(\alpha-\theta),\\
\dot{r} =v_\perp \text{cos}(\alpha-\theta),\nonumber\\
\dot{v_\parallel}=\frac{Z_i}{m_i} E_{\parallel}.\nonumber
\end{eqnarray}

In the fully kinetic version of the GTC code the perpendicular component of the velocity $(v_{\perp})$ and the gyro phase angle $(\alpha)$ can be calculated from Eq. (2) using the Boris push method \cite{Boris,Birdsall} .  In the following section we will discuss the implementation of the Boris push technique in GTC. However, for the calculation of $v_{\parallel}$ we use conventional Runge-Kutta method.

\begin{figure}[bpt]
	\centering
	\includegraphics[width=0.5\textwidth]{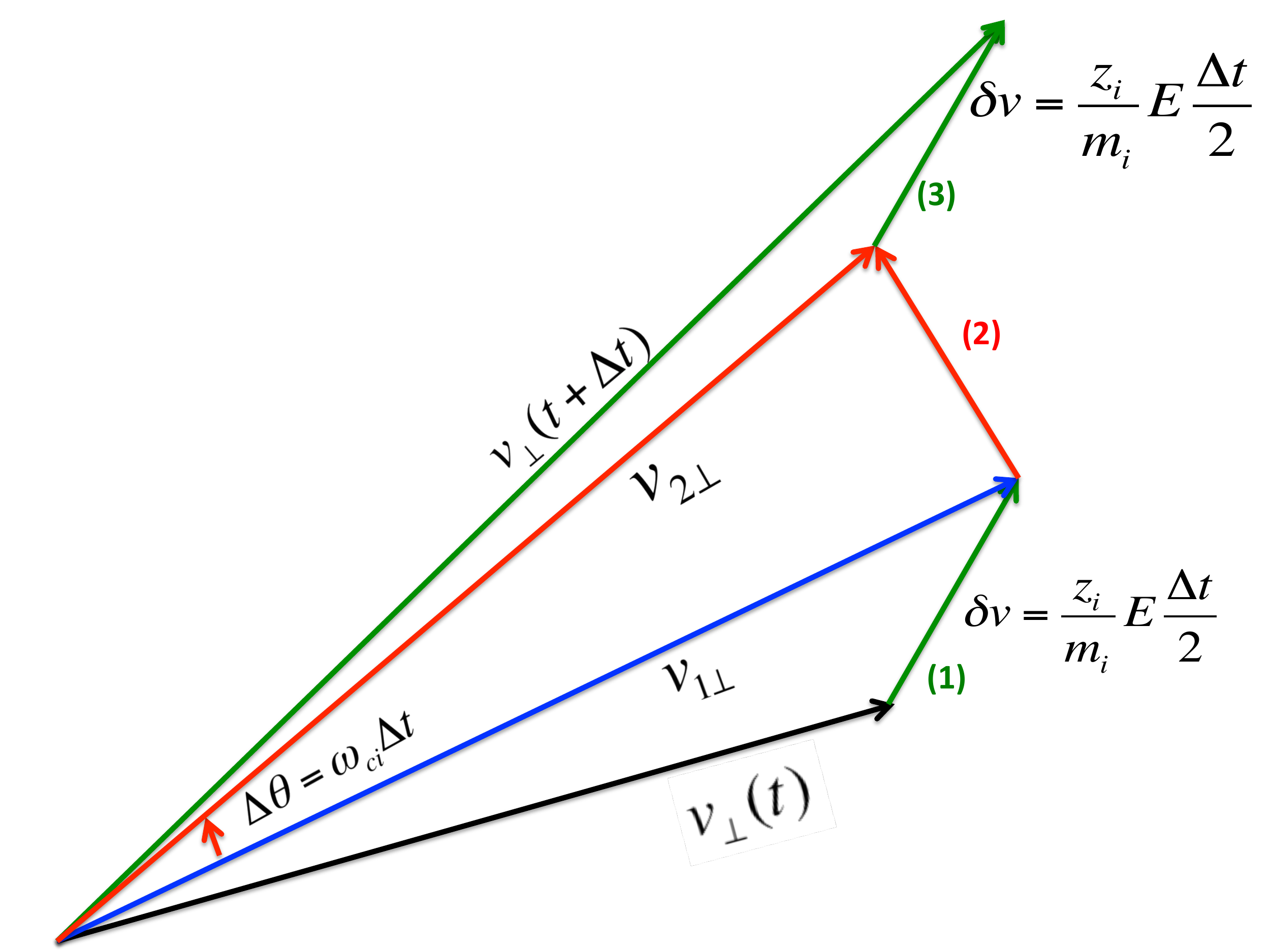}
	\caption{Schematic diagram for Boris push method. The first step indicates the addition of the first half of the electric field impulse to the velocity. The red color defines the rotation of the velocity vector in the second step. In the third step, we add the second half of the electric field impulse to the rotated velocity component.}
\end{figure}

\subsection{Boris push implementation in GTC}
The particle push is an important part of the simulation process. Eq. (2) is basically Newton's second law with the force being the Lorentz force. It is numerically challenging to integrate the particle velocity in the presence of the magnetic field. This problem can be overcome by 
defining the velocity as suggested by Boris \cite{Boris,Birdsall}. This explicit algorithm is simple to implement, with second order accuracy. It is symmetric to the time reversal, i.e., it preserves the canonical invariants\cite{penn2003}. The Boris push process can be summarized in the following three steps as described in Fig. 2. 

In the cylindrical geometry with magnetic field in the z direction, we decompose the velocity components in the direction perpendicular and parallel to the magnetic field. In the first step we add the first half of the electric field impulse to the velocity vector $\boldsymbol{v}_{\perp}(t)$ to obtain a new $\boldsymbol{v}_{1\perp}$ as

\begin{equation}
\boldsymbol{v}_{1\perp}=\boldsymbol{v}_{\perp}(t)+\delta\boldsymbol{v}_{\perp}, \quad \text{where} \quad 
\delta\boldsymbol{v}_{\perp}=\frac{Z_i}{m_i} \boldsymbol{E}_\perp \frac{\Delta t}{2}.
\end{equation}

We use (x, y) coordinates to represent $\delta\boldsymbol{v}_{\perp}$ (see Fig.1). From Eq. (2) we get
	\begin{equation}
	\left\{\begin{aligned}
	&	\delta v_x=-\frac{Z_i}{m_i}\left( \frac{\partial\phi}{\partial r}\text{cos}\theta -\frac{1}{r} \frac{\partial\phi}{\partial\theta}\text{sin}\theta\right)\frac{\Delta t}{2},\\
	&	\delta v_y=-\frac{Z_i}{m_i}\left( \frac{\partial\phi}{\partial r}\text{sin}\theta + \frac{1}{r} \frac{\partial\phi}{\partial\theta}\text{cos}\theta\right)\frac{\Delta t}{2},	
	\end{aligned}\right.
	\end{equation}
	
	and
\begin{equation}
	\left\{
	\begin{aligned}
	&	v_{1x}=v_{\perp}\text{cos}(\alpha(t)) +\delta v_x, \\
	&	v_{1y}=v_{\perp}\text{sin}(\alpha(t)) +\delta v_y. \\
	\end{aligned}
	\right.	
\end{equation}
In the second step we consider the rotation of the velocity vector $\boldsymbol{v}_{1\perp}$. The vector form of this rotation is given by 
\begin{eqnarray}
		\text{T}=\frac{Z_i}{m_i}\text{B}_0\frac{\Delta t}{2}, \nonumber\\
		u=v_{1x}+v_{1y}\text{T},
\end{eqnarray}	
and
\begin{equation}
	\left\{\begin{aligned}
	&	v_{2y}=v_{1y}-u\text{S},\\
	&	v_{2x}=u+v_{2y}\text{T},\\
	\end{aligned}\right.
\end{equation}
where $\text{S}=2\text{T}/(1+\text{T}^2)$, is also a form of rotation vector $\text{T}$ scaled to satisfy that the magnitude of the velocity should remain unchanged during the rotation. Eqs. (16) and (17) together give the rotation of the velocity vector as shown by the red color in the Fig. 2. 

In the third step we add the remaining half of the electric field impulse to the rotated vector $\boldsymbol{v}_{2\perp}$ to obtain 
\begin{equation}
\left\{
	\begin{aligned}
 v_x(t+\Delta t )=v_{2x}+\delta v_x,\\
 v_y(t+\Delta t )=v_{2y}+\delta v_y.\\
\end{aligned}
	\right. 
\end{equation}
Now we can write down the new $v_{\perp}(t+\Delta t)$ and gyro phase angle $\alpha(t+\Delta t)$ from $v_x$ and $v_y$
\begin{equation}
\left\{
	\begin{aligned}
 v_{\perp}(t+\Delta t )=\sqrt{v_x^2(t+\Delta t)+v_y^2(t+\Delta t )},\\
\text{tan}[ \alpha(t+\Delta t )]=\frac{v_y(t+\Delta t )}{v_x(t+\Delta t )},\\
\end{aligned}
	\right. 
\end{equation}
where $\alpha$ is chosen to vary in the range of $[0,2\pi]$.
\begin{figure}[bpt]
	\centering
	\includegraphics[width=0.5\textwidth]{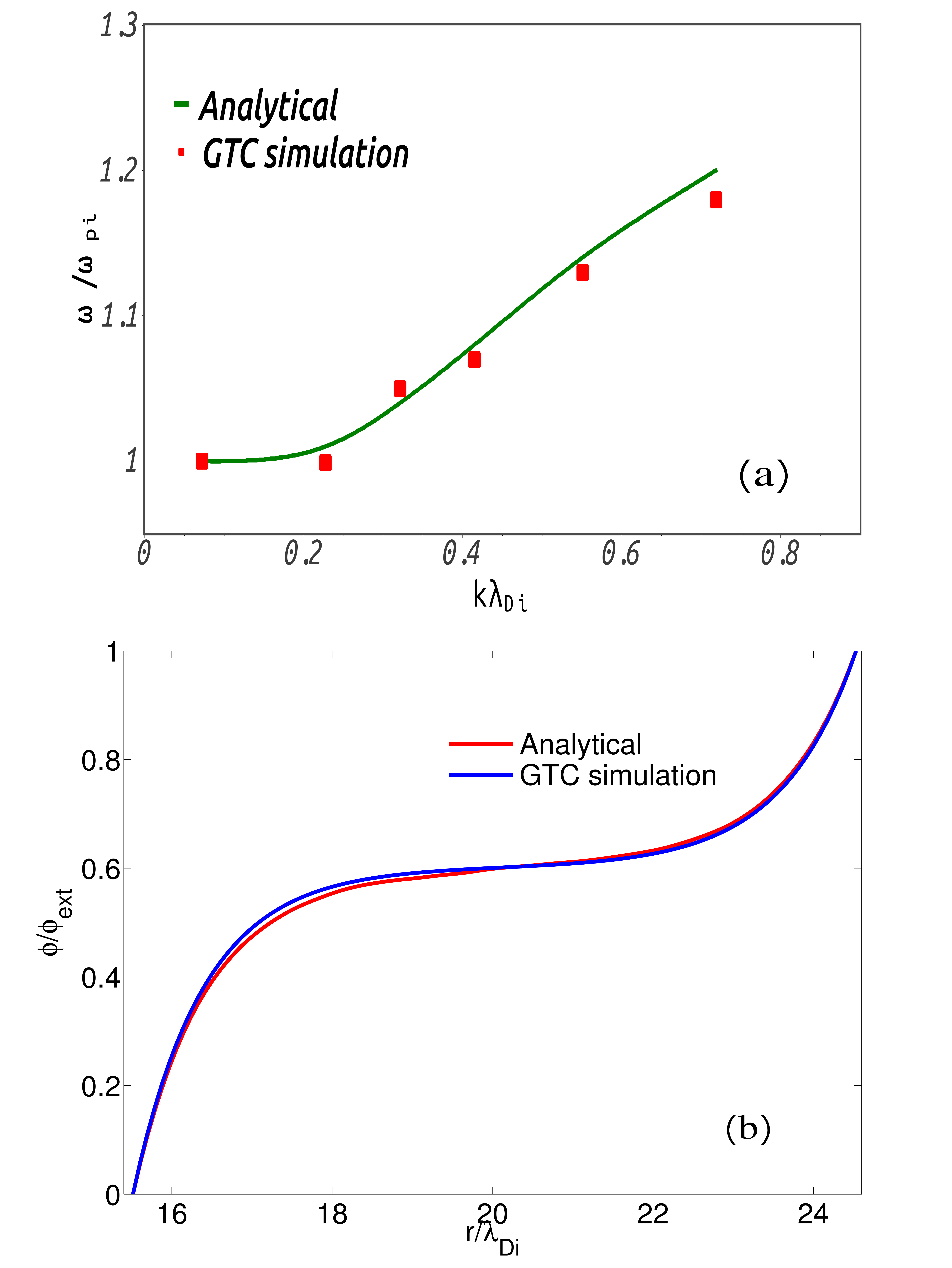}
	\caption{(a) Ion plasma oscillation frequency as a function of normalized wavelength $(k\lambda_{Di})$, and its verification with the analytical theory (cf. Eq. (22)), (b) comparison of the electrostatic potential of the ion plasma wave as a function of the normalized radius between analytical theory and GTC simulation.}
\end{figure}

\section{Verification of Normal modes}
In this section we will discuss the electrostatic normal modes with $k_{\parallel}=0$ in uniform plasmas and uniform magnetic field.
The corresponding dispersion relation can be written as
\begin{equation}
1+\chi_j=0 
\end{equation}
By considering the uniform Maxwellian background plasma using Eqs. (1) and (2), one can write down the susceptibility as\cite{Liu86}
\begin{equation}
\chi_j=-\frac{1}{k_{\perp}^2\lambda_{Dj}^2}\sum_{l=1}^{\infty}\frac{2l^2\omega_{cj}^2}{\omega^2-l^2\omega_{cj}^2}I_l(b_j)e^{-b_j},
\end{equation}
where $\lambda_{Dj}^2=\epsilon_0T_j/n_0e^2$, $b_j=k_{\perp}^2\rho_j^2/2$, $\omega_{ce}$, $\omega_{ci}$ are the electron, ion cyclotron frequencies\textcolor{red}{,} respectively\textcolor{red}{.} $\rho_e$ and $\rho_i$ are the electron and ion Larmor radius\textcolor{red}{,} respectively. There are only three electrostatic normal modes in the uniform plasma for $k_{\parallel}=0$, e.g., ion plasma oscillation, lower hybrid wave and ion Bernstein wave.

 \subsection{Ion plasma oscillation}
Unmagnetized ions and magnetized electrons support the normal mode called ion plasma oscillation when $k_{\parallel}=0$. In the massless electron limit, the ion and electron contributions to the susceptibility can be written as 
 \begin{equation}
\chi_{i}=-\frac{\omega_{pi}^2}{\omega^2}, \quad \chi_e=0
 \end{equation}
To verify the fully kinetic ion model, we carried out simulation\textcolor{red}{s} for different equilibrium plasma density (i.e., varying the ion Debye length $\lambda_{Di}$). Fig. 3(a) demonstrate that for small value of $k\lambda_{Di}$, we can recover $\omega_{pi}$, the ion plasma oscillation. In the presence of the finite ion temperature the ion plasma wave will be damped after a few oscillations because of ion Landau damping. During this process the electric field can penetrate up to the ion Debye length. GTC simulation of the ion Debye shielding effect agrees well with the analytic theory [Fig. 3(b)]. In the simulations the boundary conditions for the electrostatic potential are $\phi=0$ at the inner boundary and $\phi=$constant at the outer boundary.  These one-dimensional simulations are carried out using the full-$f$ method. The system length is about 10 ion Debye lengths. The number of grid points in radial, poloidal, and parallel direction is Nx=100, Ny=100, and Nz=32, respectively. A total of 4000 particles per cell are used. Initially the particles are loaded uniformly with a Maxwellian velocity distribution. The initial fluctuations are due to the random noise.

\subsection{Lower Hybrid waves}
\begin{figure}[bpt]
	\centering
	\includegraphics[width=0.5\textwidth]{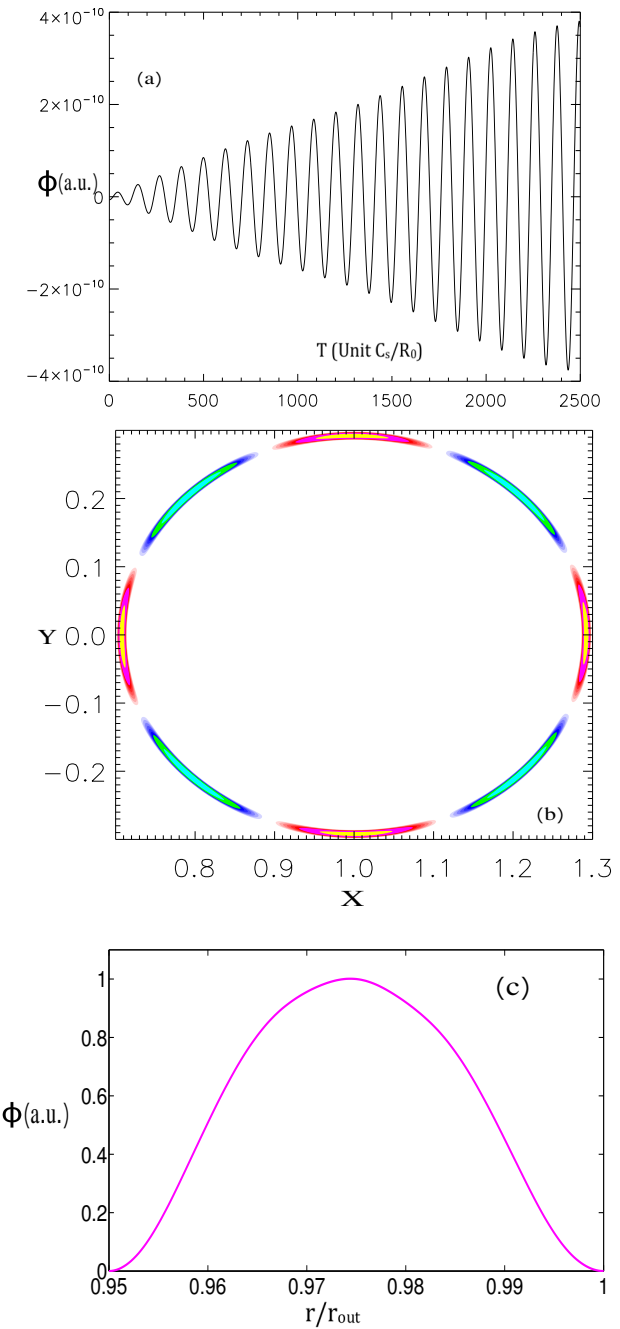}
	\caption{(a) Time history of m=4 lower hybrid wave amplitude excited by the antenna (b) Poloidal mode structure of electrostatic potential $\phi$ and (c) Radial profile of $\phi$.}
\end{figure}

\begin{figure}[bpt]
	\centering
	\includegraphics[width=0.5\textwidth]{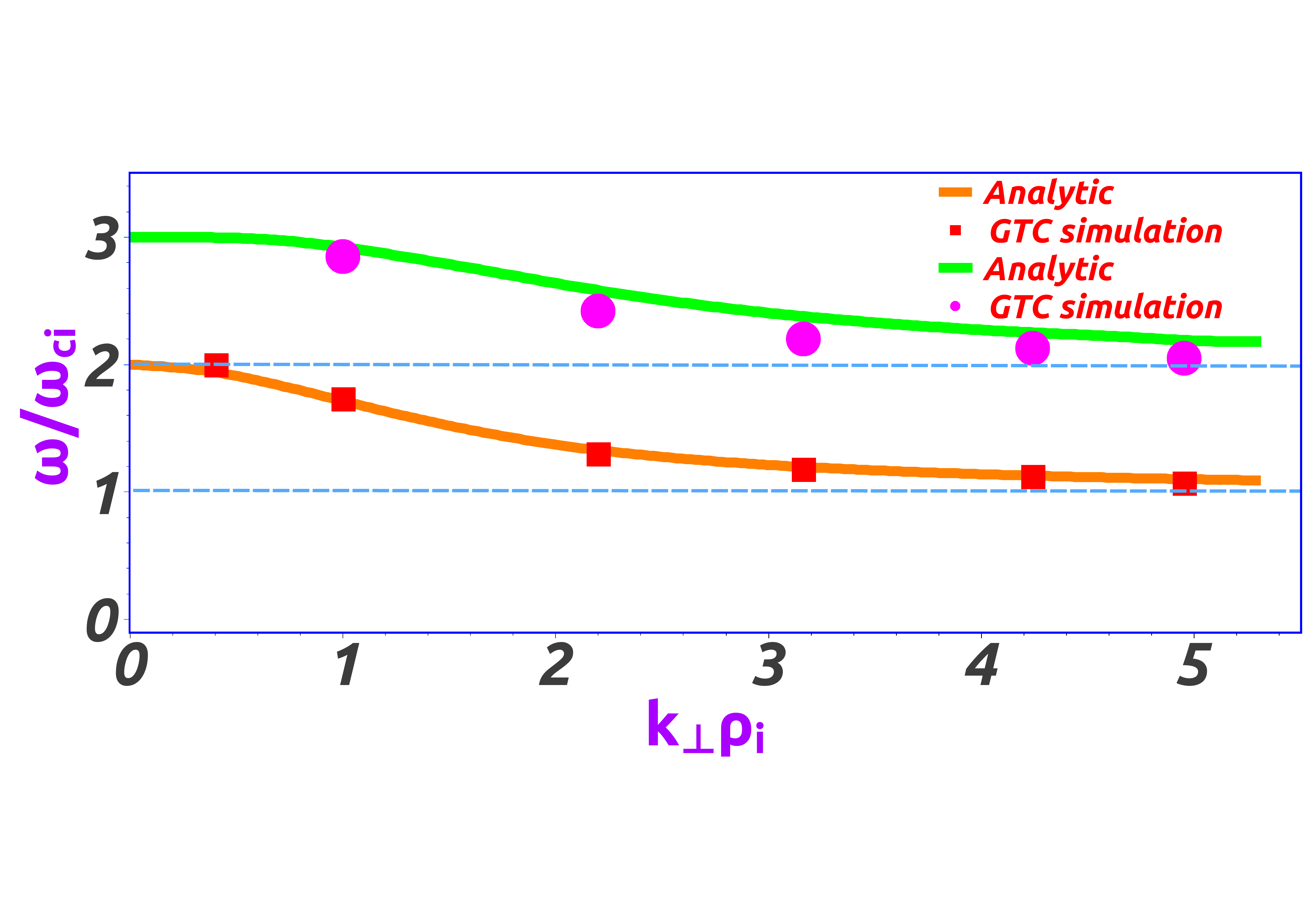}
	\caption{Comparison of ion Bernstein wave dispersion relation between the analytical solution of Eq.(27) and the GTC simulations with for the first and second harmonics.}
\end{figure}
Lower hybrid waves are space-charge waves in the  frequency range $\omega_{ci}\ll\omega \ll \omega_{ce}$. In this limit ion motion can be taken to be unmagnetized and the ion susceptibility become\textcolor{red}{s}\cite{Liu86}
 \begin{equation}
\chi_{i}=-\frac{\omega_{pi}^2}{\omega^2-\omega_{ci}^2}\simeq-\frac{\omega_{pi}^2}{\omega^2}
 \end{equation}
Now we consider the finite mass of the electron. For such normal modes in the magnetized plasma with $k_{\perp}\rho_e\ll 1$, $\chi_e$ is dominated by $l=1$ term as
\begin{equation}
\chi_{e}=\frac{\omega_{pe}^2}{\omega_{ce}^2},
\end{equation}
which arises due to the guiding center polarization drift.  We implement the electron polarization term in GTC similar to the ion polarization term calculated in the gyrokinetic simulation.

By using Eq. (20) in the limit of $\omega_{pe}\gg\omega_{ce}$, the frequency of the lower hybrid wave is 
\begin{equation}
\omega_{LH}^2=\frac{\omega_{pi}^2}{(1+\omega_{pe}^2/\omega_{ce}^2)}\approx {\omega_{ci}\omega_{ce}}
\end{equation}

We use an artificial antenna to excite these modes and to verify the mode structure and frequency in our simulation. The antenna is implemented for the electrostatic potential $\phi$ as follows \cite{huassen2010}

\begin{equation}
\phi_{ant}=\hat{\phi}(r)\text{sin}(n_{ant}\zeta-m_{ant}\theta)\text{sin}(\omega_{ant}t)
\end{equation}

To find the eigenmode frequency of the system, we carry out the scan with different antenna frequencies and find out the frequency in which the mode has the maximum growth of the amplitude. That frequency is then identified as the eigenmode frequency of the system.

In our simulation the background plasma density is uniform with a uniform temperature. The simulations are all linear and electrostatic. We apply a poloidal mode filter to select only the $m = 4$ mode. In this simulation $\omega_{pe}=3.4\omega_{ce}$, $\omega_{pi}=145.2\omega_{ci}$ and $m_e/m_i=5.44618\times 10^{-4}$. Fig. 4(a) is the time evolution of the (m=4) LHW excited with an antenna frequency $\omega_{ant}=41.1\omega_{ci}$, which gives the maximal growth of the wave amplitude. Fig. 4(b) is the poloidal mode structure of the electrostatic potential.  The simulation result of the LHW frequency $\omega_{LH}=41.1\omega_{ci}$ agrees well with the analytical result of $42.8\omega_{ci}$  (cf. Eq. (25)).

\subsection{Ion Bernstein waves}
An important kinetic feature for the normal modes of magnetized ion plasma is the finite Larmor radius effect, which modifies the cold plasma mode with frequency close to the harmonics of ion cyclotron frequency, known as ion Bernstein waves (IBW).  Using Eq. (20) the dispersion relation of the IBW becomes
\begin{equation}
1+\frac{\omega_{pe}^2}{\omega_{ce}^2}=\frac{1}{k_{\perp}^2\lambda_{Di}^2}\sum_{l=1}^{\infty}\frac{2l^2\omega_{ci}^2}{\omega^2-l^2\omega_{ci}^2}I_l(b_i)e^{-b_i}
\end{equation}

Fig. (5) shows the dispersion relation of the ion Bernstein wave obtained by solving the Eq. (27) analytically with $\omega_{pi}=10\omega_{ci}$ and $\omega_{pe}=0.234\omega_{ce}$ for the first and second harmonics. To compare our GTC simulation with analytical results we carried out our simulations in different wavelengths and for different harmonics $l=1$ and $l=2$. Fig. (5) demonstrates a good agreement between the analytical and  GTC simulation results of the IBW frequency. These simulations are carried out using the $\delta f$ method. The number of grid points in radial, poloidal, and parallel direction is Nx=100, Ny=200, and Nz=32, respectively.   A total of 90 particles per cell are used.  We have carried out the convergence study of the real frequency as a function of the number of particles per cell (cf. Fig. 6). The simulation results do not depend sensitively on the number of particles, as the grid numbers per wavelength is sufficiently large (100 in this case). In our simulation we have $\omega \Delta t < 0.01$, where $\omega$ is the frequency of the normal mode and $\Delta t$ is the time step. So, we have more than 600 time steps per wave period. To measure the wave frequency we count the number of time steps in several wave periods from the time history of the wave amplitude. The uncertainty in measuring the frequency is defined as the inverse of number of time steps. This provides a better accuracy than the FFT in measuring the frequency.

\section{Discussions}
In summary, with the implementation of the fully kinetic ion and drift kinetic electron model, GTC is particularly applicable to problems in which the electrostatic normal mode frequency ranges from ion Bernstein wave to  lower hybrid waves. This new simulation model should have wide applications in the areas of radio frequency heating and current drive, control of MHD instability, and other nonlinear phenomenon. The model is more efficient for the physical process with $\omega\ll\omega_{ce}$, $k_{\parallel}\ll k_{\perp}$, and can handle the realistic electron-to-ion mass ratio, by removing the fast electron gyro motion from the wave dynamics. The LHW and IBW excitation by artificial antenna provides the verification of the mode structure, and the frequency using the predicted by linear theory. Our initial verification suggests that the present simulation model is promising and can incorporate a broad range of realistic issues  (toroidal geometry, electromagnetic effects, nonlinear kinetic effects, nonlinear ion Landau damping, parametric instabilities,  and ponderomotive effects).

\begin{figure}[bpt]
	\centering
	\includegraphics[width=0.5\textwidth]{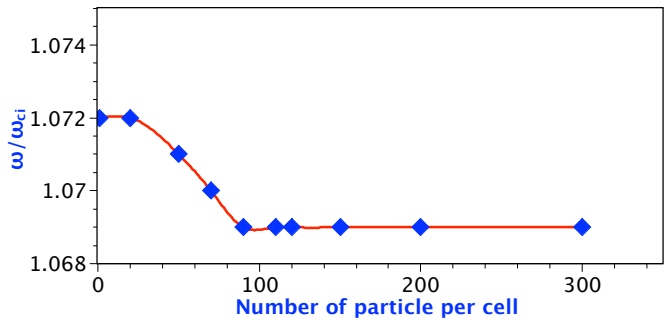}
	\caption{Convergence study of real frequency as a function of number of particles per cell. Red line represents the fit to the data points. }
\end{figure}

 \begin{acknowledgments}
 This work is supported by the Trialpha Energy Inc., U.~S.~Department of Energy (DOE), and China National Magnetic Confinement Fusion Science Program, Grant No. 2013GB111000. Simulations were performed using supercomputers at ORNL, NERSC and NSCC-TJ.
 \end{acknowledgments}

\bibliographystyle{apsrev4-1}
\bibliography{FKGTC}

\end{document}